# Satellite Observations of Separator Line Geometry of Three-Dimensional Magnetic Reconnection

C. J. Xiao[1*], X. G. Wang[2,3], Z. Y. Pu[4*], Z. W. Ma[5,6], H. Zhao[1], G. P. Zhou[1], J. X. Wang[1], M. G. Kivelson[7], S. Y. Fu[4], Z. X. Liu[8], Q. G. Zong[9], M. W. Dunlop[10], K-H. Glassmeier[11], E. Lucek[12], H. Reme[13], I. Dandouras[13], C. P. Escoubet[14]

1. *National Astronomical Observatories, Chinese Academy of Sciences, Beijing 100012, China,*

2. *School of Physics, Peking University, Beijing 100871, China,*

3. *State Key Lab of Materials Modification by Beams, Dalian University of Technology, Dalian 116024, China*

4. *School of Earth and Space Sciences, Peking University, Beijing 100871, China,*

5. *Institute for Fusion Theory and Simulation, Zhejiang University, Hangzhou 310027, China*

6. *Institute of Plasma Physics, Chinese Academy of Sciences, Hefei 230031, China,*

7. *Institute of Geophysics and Planetary Physics, University of California, Los Angeles, California, USA,*

8. *CSSAR, Chinese Academy of Sciences, Beijing 100080, China,*

9. *Center for Atmospheric Research, University of Massachusetts Lowell, Lowell, MA 01854-3629, USA*

10. *Space Sciences Division, SSTD, Rutherford Appleton Laboratory, Chilton, Oxfordshire, OX11 0QX, UK*

11. *IGEP, Technische Universität Braunschweig, Braunschweig, Germany,*

12. *Blackett Laboratory, Imperial College, London, UK*

13. *Centre d'Etude Spatiale des Rayonnements, BP 4346, 31028 Toulouse Cedex 4, France ,*

14. *ESA/ESTEC, Postbus 299, 2200 AG Noordwijk, The Netherlands*

(e-mails: cjxiao@ourstar.bao.ac.cn, xgwang@dlut.edu.cn, or zypu@pku.edu.cn).

Detection of a separator line that connects magnetic nulls and the determination of the dynamics and plasma environment of such a structure can improve our understanding of the three-dimensional (3D) magnetic reconnection process [1-9]. However, this type of field and particle configuration has not been directly observed in space plasmas. Here we report the identification of a pair of nulls, the null-null line that connects them, and associated fans and spines in the magnetotail of Earth using data from the four Cluster spacecraft. With $d_i$ and $d_e$ designating the ion and electron inertial lengths, respectively, the separation between the nulls is found to be ~ 0.7±0.3 $d_i$ and an associated oscillation is identified as a lower hybrid wave with wavelength ~ $d_e$. This *in situ* evidence of the full 3D reconnection geometry and associated dynamics provides an important step toward to establishing an observational framework of 3D reconnection.

In general, 3D magnetic reconnection occurs on a separator line that is analogous to a two-dimensional (2D) X-line. The legs of this line, called separatrices, correspond to fans ($\Sigma$-surfaces) bounded by spines ($\gamma$-lines) that emerge from the nulls [1-4]. 3D reconnection models are fundamentally based on chains of A-B null pairs [1-4, 8-9]. [The classification of a null depends on whether the field along the separator line converges (A) or diverges (B) from the null point, as shown in Figure 1d.] Identification of A-B null pairs and identification of the separator and associated structures such as spines and fans is a fundamental to increasing our understanding of 3D reconnection.

Identification of null-null lines and their neighbouring 3D structure and associated dynamics has been attempted for solar coronal plasmas [10-11]. Recently, Xiao et al. successfully identified an isolated magnetic null point related to a reconnection region in the magnetotail of the earth [9]. These authors made use of measurements provided by the Cluster mission [12], which provides data from four similarly instrumented spacecraft that form a tetrahedron in space, thereby providing unique opportunities to detect small-scale 3D plasma structures. In their analysis, these authors applied the Poincaré-Index method [13-14] to infer the presence of a true magnetic null point.

Here we analyze an event on October 1, 2001 between 09:36 UT and 09:55 UT during which time the Cluster spacecraft meandered several times around a reconnection region in the magnetotail, near ~ (-16.3, 7.9, 0.9) $R_E$ GSM (in geocentric solar magnetospheric coordinates). As shown in Figure 1a, the four spacecraft were separated by less than 2250 km. Four second average data from the Fluxgate Magnetometer (FGM) [15] for the magnetic field ($B$) and the Cluster Ion Spectrometer (CIS) [16], for the plasma density ($N$), and velocity ($V$) during the interval 09:45-09:51 UT are plotted for spacecraft C4 in Fig. 2a-2c. The event has previously been extensively studied assuming a two-dimensional structure for reconnection [17-20]. A tailward passage of the X-line with respect to the Cluster tetrahedron from 09:47-09:51UT is shown. The measured magnetic field and flow velocity patterns match the 2D Hall reconnection picture [21-22].

Vector field theory [23] implies that X-lines are structurally unstable and that any small perturbation would break them apart into pairs of nulls [2]. We, therefore, considered that this X-line passage event could be a potential candidate for detecting a pair of nulls. In the interval 09:48:20-09:48:50 UT, the Cluster tetrahedron encountered the "X-line" as indicated by the observed flow reversal and the quadrupolar Hall magnetic field perturbations. In order to identify a null pair in the reconnection region during the period under consideration we apply the Poincaré Index method [13-14], which has been successfully used in previous work for identifying magnetic nulls from its change of sign [9]. For this purpose we use high resolution (0.04s) magnetic field data from all four spacecraft and find that the Poincaré Index jumps up first to +1, then jumps down to -1, and returns afterwards to +1 again (see Fig. 2d-2g). We may, therefore, conclude that some magnetic nulls were present within the Cluster tetrahedron. We can also verify the result by linear interpolation as shown in Fig. 1a.

The total relative error in calculating $\delta \mathbf{B}$ can be estimated from the ratio $|\nabla \cdot \mathbf{B}|/|\nabla \times \mathbf{B}|$ [9, 24-25]. In this event $|\nabla \cdot \mathbf{B}|/|\nabla \times \mathbf{B}|$ is a few hundredths at the times when the Poincaré Index is $\pm 1$. The types of these nulls can be identified from three eigenvalues

of the corresponding $\delta\mathbf{B}$ matrix around the nulls. If the real parts of two eigenvalues for a null are negative, we call it a negative or A-type. Otherwise, we call it a positive or B-type [1-4]. We find that the Cluster tetrahedron first encounters an A-null in the interval 09:48:24.166 - 09:48:25.682UT, then a B-null in the interval 09:48:26.975 - 09:48:29.830UT. Characteristics of the null pair observed are shown in Table 1.

The detailed analysis of the null-pair characteristics is consistent with a 3D reconnection configuration such as that plotted in Fig. 1d, where the A-B null line is analogous to the X-line in 2D model (as illustrated in Fig. 1c), with the legs replaced by the fan planes $\Sigma_A$ and $\Sigma_B$ that are bounded by the spines $\gamma_B$ and $\gamma_A$ respectively [1-4]. The surfaces $\Sigma_A$ and $\Sigma_B$ are determined by the eigenvectors $\mathbf{a}_2$ and $\mathbf{a}_3$, and $\mathbf{b}_2$ and $\mathbf{b}_3$ in Table 1 respectively; while the spines $\gamma_A$ and $\gamma_B$ are determined by the eigenvectors $\mathbf{a}_1$ and $\mathbf{b}_1$. The field lines above the upper halves of the $\Sigma$ surfaces and those below the lower halves merge and cross-link at the null-null line.

Theory predicts that A and B nulls that connect to form a 3D reconnection geometry, obey following features [2] : 1) $\mathbf{a}_3$ corresponding to the smaller negative eigenvalue of the A-null and $\mathbf{b}_3$ corresponding to the smaller positive eigenvalue of the B-null form an A-B null line. 2) The null-null line is the intersection of the two fan planes $\Sigma_A$ and $\Sigma_B$. Therefore the null-null line ($\mathbf{b}_3$) and $\gamma_B$ ($\mathbf{b}_1$) as the boundary of $\Sigma_A$ should be on the same surface $\Sigma_A$. 3) Similarly, both the $\gamma_A$ ($\mathbf{a}_1$) and the null-null line $\mathbf{a}_3$ should be on the surface $\Sigma_B$. In other words, if the $\Sigma$ surfaces are approximately planar, the angles between $\gamma_A$ ($\mathbf{a}_1$) and $\Sigma_B$, and between $\mathbf{a}_3$ and $\Sigma_B$ should be small.

Indeed, the observation shows that in this event: 1) $\mathbf{a}_3$ and $\mathbf{b}_3$ approximately coincide creating the A-B line with an angle of ~10° between them. 2) As shown in Table 1, $\gamma_B$ ($\mathbf{b}_1$) and $\mathbf{b}_3$ are approximately located on $\Sigma_A$ with inclinations less than 4° and 2°, respectively. 3) $\gamma_A$ ($\mathbf{a}_1$) and $\mathbf{a}_3$ are approximately on $\Sigma_B$ with inclinations less than 2° and 0°, respectively. We thus conclude that the data meet the criteria of 3D reconnection, and the Cluster spacecraft encountered an A-B null pair.

The null-null line is found almost to align with the y direction as shown in Fig. 1d, and the angles between the separatrices (fans, or legs of the "X-point") calculated from the data are ~ 42° - 46°, as shown in Table 1, This is also in reasonable agreement with the result estimated from 2D electric potential patterns [18]. Therefore the x-z plane projection of the 3D reconnection configuration observed is very similar to the previously reported 2D "X-line" reconnection geometry of this event [17-18].

The length of the null-null line cannot be measured precisely. However, the time interval from the Cluster tetrahedron's first encounter with the A-null at 09:48:24.166 UT to its first encounter to the B-null at 09:48:26.974 UT is 2.808 seconds. Since the time resolution of the velocity data is only 4s, we must apply other means to calculate the drift speed of the magnetic configuration relative to the Cluster tetrahedron. An appropriate approach is to calculate the velocity based on the null positions at adjacent moments. This can be done by locating successive null positions in the Cluster frame using linear interpolation [13] and then calculating the average drift speed and its mean square error to obtain 310±120 km/s. Multiplying the drift velocity by the above separation time, we find for the approximate null-null separation 860±340 km≈ 0.7±0.3 $d_i$, where $d_i$ =1320 km is the ion inertial length for the ion density of 0.03 cm$^{-3}$ shown in Fig. 2a.

In examining the high resolution data to characterize the null trajectory we also detected oscillations in the lower hybrid frequency range of 13Hz (see Figure 3b). Lower hybrid oscillations have been reported in conjunction with reconnection in other observations [26-27], as well as in experiments and simulations [5-7, 28]. In this event, the oscillation is found on the separator line. The polarization of the wave is found in the L-M plane, where L and M are the directions of maximum and intermediate variances of magnetic field [25], with the preferable direction almost along the line diagonal to the L and M axes. The wavelength can be calculated from the trajectory and drift velocity of the null as 24±10 km ~ $d_e$ =30 km. From cross-wavelet analysis of the magnetic field around the null-null line [29-30], one finds that the wave vector is almost perpendicular to the magnetic field with $k_\parallel / k_\perp < 0.1$. This suggests that the dynamics in the neighbourhood of the null-null line is characterized by lower hybrid waves in the magnetized electron-unmagnetized ion regime. From the fact that $1/k_\parallel$ is at least an order of magnitude larger than $d_e$, lying in the range of the null-null separation, it is likely that the null-pair is related to the lower hybrid perturbation. The observed lower-hybrid wave in the vicinity of the null-null line is consistent with the density dip at the X-point reported in observations and simulations[21-22].

**Acknowledgements** This work is supported by the NSFC Programs (Grant No. 40390150, 40504021, 10233050, 10575018, 40536030, 40425004, and 40228006) and the National Basic Research Program of China (Grant No.2006CB806300), as well as the CAS Project KJCX2-YW-T04 and the China Double Star-Cluster Science Team. C. J. also thanks Aslak Grinsted at University of Lapland to support the MatLab wavelet coherence package.

C. J., X. G., and Z. Y. are first authors with equal contributions to theoretical and data analysis. Z. W. has also participated in the analysis, and Z. H, G. P. and J. X. have developed analysis tools. Other co-authors have provided the Cluster observation data and involved in discussions. M. G. and Z. W. have also contributed to the final revision of the paper.

**Competing interests statement** The authors declare that they have no competing financial interests.

**Correspondence** and requests for materials should be addressed to Z.Y., X.G. or C.J. (e-mails: cjxiao@ourstar.bao.ac.cn, xgwang@dlut.edu.cn, or zypu@pku.edu.cn).

Figure 1: The 3D reconnection configuration. (a) The locations of the Cluster tetrahedron at two times separated by 2.808 seconds, in a frame moving with the average drift speed of nulls and origin at the 09:48:25.593 UT position of spacecraft C3, (-16.2, 7.9, 0.5) $R_E$ in GSM coordinates. The yellow dots indicate the locations of two nulls calculated from linear interpolation. (b) The reconnection region in the magnetotail (red square). (c) The x-z plane 2D projection of the 3D reconnection geometry. (d) Topological structure of 3D reconnection with observed A and B-nulls, fan surfaces $\Sigma_A$ and $\Sigma_B$ and spine lines $\gamma_A$ and $\gamma_B$ in GSM. The fans $\Sigma_A$ (bounded by $\gamma_B$) and $\Sigma_B$ (bounded by $\gamma_A$) intersect along the null-null line (shown in purple). The characteristic directions of all vectors and surfaces were calculated from Cluster data (Table 1).

Figure 2: The Cluster measurements and Poincaré index calculation in the reconnection event on 1 October 2001. (a)-(c): Overview of the 4-s resolution data of the plasma density (**N**), magnetic field (**B**), and velocity (**V**) for C4 from 09:45 to 09:51 UT, where the $V_x > 0$ intervals are masked red, and the $V_x < 0$ interval is masked yellow. The blue masked area is the flow reversal interval where the nulls are detected; (d)-(f): The intercalibrated high resolution (0.04-s) magnetic field data for C1-C4 for a period in which nulls were identified; and (g): The corresponding Poincaré Index calculated during 09:48:20-09:48:50 UT.

Figure 3: The null's drift velocity and the lower-hybrid oscillation signatures. (a) The velocity of the null drifting inside the cluster tetrahedron in LMN coordinates with L=(-0.22, 0.97, -0.05), M=(0.97, 0.23, 0.11), N=( 0.12, -0.03, -0.99) in GSM. The L-axis is almost aligned with the null-null line shown in Table 1. (b) The FFT spectrum of the null point oscillation shows maximum power at 10-15Hz, consistent with the lower hybrid frequency ($\omega_{LH} = \sqrt{\Omega_i \Omega_e} = 13.0 H_z$, with magnetic field 20 nT). (c) Hodogram of the velocity of the null drifting in the L-M plane starting at the blue spot; (d) The wave vector in LMN coordinates calculated by cross wavelet analysis shows that $k_N > k_M >> k_L$ in the frequency range of 10-13 Hz (a.u. is the abbreviation of arbitrary unit).

**Table 1: Characteristics of the null pair observed on the 1 October, 2001.**

| Time | 09:48:25.593 UT | 09:48:28.447 UT |
|---|---|---|
| Poincaré Index | 1 | -1 |
| Null Type | A-null (Negative) | B-null (Positive) |
| $\delta \mathbf{B}$ | $\begin{bmatrix} 0.0020 & -0.00066 & 0.019 \\ -0.0027 & -0.00045 & -0.0031 \\ 0.0027 & -0.00065 & -0.0015 \end{bmatrix}$ | $\begin{bmatrix} 0.0011 & 0.0015 & 0.018 \\ -0.0023 & 0.0015 & -0.0016 \\ 0.0034 & -0.0000067 & -0.0017 \end{bmatrix}$ |
| $\nabla \cdot \mathbf{B}$ | $2.2 \times 10^{-5}$ | $8.6 \times 10^{-4}$ |
| $\frac{|\nabla \cdot \mathbf{B}|}{|\nabla \times \mathbf{B}|}$ | 0.13% | 1.1% |
| Eigenvalues | $\lambda_1 = +0.0081$ <br> $\lambda_2 = -0.0070$ <br> $\lambda_3 = -0.0011$ | $\lambda_1 = -0.0082$ <br> $\lambda_2 = +0.0072$ <br> $\lambda_3 = +0.0018$ |
| Eigenvectors | $\mathbf{a}_1 = (-0.89, 0.38, -0.27)$ <br> $\mathbf{a}_2 = (0.89, 0.17, -0.42)$ <br> $\mathbf{a}_3 = (-0.24, 0.97, -0.01)$ | $\mathbf{b}_1 = (0.88, 0.13, -0.46)$ <br> $\mathbf{b}_2 = (0.84, -0.43, 0.32)$ <br> $\mathbf{b}_3 = (-0.09, 0.99, -0.09)$ |
| Angles between | $\gamma_A$ ($\mathbf{a}_1$) and $\Sigma_B$ ($\mathbf{b}_2, \mathbf{b}_3$) → 2°;   $\gamma_A$ ($\mathbf{a}_1$) and $\Sigma_A$ ($\mathbf{a}_2, \mathbf{a}_3$) → 42° <br> $\gamma_B$ ($\mathbf{b}_1$) and $\Sigma_A$ ($\mathbf{a}_2, \mathbf{a}_3$) → 4°;   $\gamma_B$ ($\mathbf{b}_1$) and $\Sigma_B$ ($\mathbf{b}_2, \mathbf{b}_3$) → 46° <br> null-null line $\mathbf{a}_3$ and $\Sigma_B$ ($\mathbf{b}_2, \mathbf{b}_3$) → 0° <br> null-null line $\mathbf{b}_3$ and $\Sigma_A$ ($\mathbf{a}_2, \mathbf{a}_3$) → 2° | |